\documentclass[aps,reprint,twocolumn,citeautoscript,showpacs,amsmath,amssymb,prl]{revtex4-1}
\usepackage{graphicx}% Include figure files
\usepackage{dcolumn}% Align table columns on decimal point
\usepackage{bm}% bold math
\usepackage{times}% font that prx use
\usepackage{color}
\begin{document}
\title{Effects of transition metal substitutions on the incommensurability and spin fluctuations in BaFe$_2$As$_2$ by elastic and inelastic neutron scattering}
\author{M. G. Kim,$^{1,2}$ J. Lamsal,$^{1,2}$ T. W. Heitmann,$^{3}$ G. S. Tucker,$^{1,2}$ D. K. Pratt,$^{1,2}$ S. N. Khan,$^{1,4}$ Y. B. Lee,$^{1,2}$ A. Alam,$^{1}$ A. Thaler,$^{1,2}$ N. Ni,$^{1,2}$ S. Ran,$^{1,2}$ S. L. Bud'ko,$^{1,2}$ K. J. Marty,$^5$ M. D. Lumsden,$^5$ P. C. Canfield,$^{1,2}$ B. N. Harmon,$^{1,2}$ D. D. Johnson,$^{1,6}$ A. Kreyssig,$^{1,2}$ R. J. McQueeney,$^{1,2}$ A. I. Goldman$^{1,2,}$}\email{goldman@ameslab.gov}
\affiliation{$^1$Ames Laboratory, U.S. DOE, Iowa State University, Ames, IA 50011, USA}
\affiliation{$^2$Department of Physics and Astronomy, Iowa State University, Ames, IA 50011, USA}
\affiliation{$^3$The Missouri Research Reactor, University of Missouri, Columbia, MO 65211, USA}
\affiliation{$^4$Department of Physics, University of Illinois, Urbana, IL 61801, USA}
\affiliation{$^5$Quantum Condensed Matter Division, Oak Ridge National Laboratory, Oak Ridge, TN 37831, USA}
\affiliation{$^6$Department of Materials Science \& Engineering, Iowa State University, Ames, IA 50011, USA}

\date{\today}

\begin{abstract}
The spin fluctuation spectra from nonsuperconducting Cu-substituted, and superconducting Co-substituted, BaFe$_2$As$_2$ are compared quantitatively by inelastic neutron scattering measurements and are found to be indistinguishable. Whereas diffraction studies show the appearance of incommensurate spin-density wave order in Co and Ni substituted samples, the magnetic phase diagram for Cu substitution does not display incommensurate order, demonstrating that simple electron counting based on rigid-band concepts is invalid.  These results, supported by theoretical calculations, suggest that substitutional impurity effects in the Fe plane play a significant role in controlling magnetism and the appearance of superconductivity, with Cu distinguished by enhanced impurity scattering and split-band behavior.
\end{abstract}

\maketitle

{\par}The role of chemical substitution and its effects on structure, magnetism and superconductivity have become central issues in studies of the iron-pnictide superconductors.\cite{Johnston_2010,CandB,PandG,Stewart_2011} This is particularly true for transition-metal ($M$) substitution on  Fe sites, resulting, nominally,  in electron doping of the FeAs layers. When low concentrations of Co,\cite{sefat_superconductivity_2008,ni_effects_2008} Ni,\cite{li_superconductivity_2009,canfield_2009} Rh,\cite{ni_phase_2009,han_2009} Pt\cite{saha_superconductivity_2010} and Pd\cite{ni_phase_2009,han_2009} replace Fe, the structural transition temperature ($T_S$) and the antiferromagnetic (AFM) transition temperature ($T_N$) are both suppressed to lower values and split with $T_S$ $>$ $T_N$.\cite{sefat_superconductivity_2008,ni_effects_2008,lester_2009,li_superconductivity_2009,ni_phase_2009,kreyssig_suppression_2010,wang_electron-doping_2010}
When the structural and magnetic transitions are suppressed to sufficiently low temperatures, superconductivity emerges below $T_c$ and coexists with antiferromagnetism over some range of concentration. Moreover, for Co, Rh and Ni substitutions in BaFe$_2$As$_2$, neutron diffraction measurements manifest a distinct suppression of the magnetic order parameter in the superconducting regime $(T < T_c$), which clearly indicates competition between AFM order and superconductivity.\cite{pratt_coexistence_2009,christianson_2009,kreyssig_suppression_2010,wang_electron-doping_2010,fernandes_unconventional_2010}

{\par}Cu substitution in BaFe$_2$As$_2$, in contrast, suppresses the magnetic and structural transitions, but does not support superconductivity\cite{canfield_2009,CandB} except, perhaps, below 2~K over a very narrow range in composition.\cite{ni_cu_2010}   This dichotomy between Co and Ni substitutions and that of Cu is also realized in quaternary fluoroarsenides.\cite{Matsuishi_2009} However, for Co/Cu co-substitutions in BaFe$_2$As$_2$, at a fixed non-superconducting Co concentration, the addition of Cu first \emph{promotes} and then \emph{suppresses} $T_c$.\cite{ni_cu_2010}  It has been suggested that previously neglected impurity effects play an important role in this behavior.\cite{canfield_2009,fernandes_2012}  The effects of impurity scattering are also neglected in a simple rigid-band picture for $M$ substitutions, which, at least for Co substitution in BaFe$_2$As$_2$, seems to adequately account for the evolution of angle-resolved photoemission spectroscopy (ARPES)\cite{CLiu_2011}, Hall effect, and thermoelectric power (TEP) measurements with concentration.\cite{Mun_2009} The rigid-band model has also been used successfully to model the suppression of the AFM transition temperature and ordered moment in Ba(Fe$_{1-x}$Co$_x$)$_2$As$_2$ for ``underdoped" samples.\cite{fernandes_unconventional_2010}  Nevertheless, this approach now faces strong challenges from recent theoretical and experimental studies.\cite{Wadati_2010,Bittar_2011,Levy_2012} Further comparative studies of Co, Ni and Cu substitutions are needed and may provide clues regarding both the nature of unconventional superconductivity in the iron pnictides and clarify the effects of $M$ substitutions.

{\par}Because a strong link between superconductivity and spin fluctuations in  iron pnictides is generally acknowledged,\cite{Johnston_2010,PandG,Stewart_2011}  it is important to  establish first whether there are any differences between the spin fluctuation spectra between superconducting (e.g., Co) and nonsuperconducting (e.g., Cu) substituted samples. Here we report on single crystal inelastic scattering measurements of the spin fluctuation spectra of Co and Cu substituted samples, with similar suppressions of the magnetic and structural ordering temperatures relative to the parent BaFe$_2$As$_2$ compound. We show that there are no quantitative differences in the normal state spin fluctuation spectra.  Therefore, we must search elsewhere for evidence of differences between Co/Ni and Cu substitutions in relation to their superconducting properties. To this end, we performed single crystal neutron diffraction measurements of the magnetic ordering in Ba(Fe$_{1-x}M_x$)$_2$As$_2$  with $M$ either Ni or Cu.  Observations of incommensurate spin-density-wave order, in particular, are a very sensitive probe of the nature of Fermi-surface nesting in the iron pnictides and, therefore, may be used to study impurity effects as a function of the $M$ doping. We find that, like the Co-substituted compound,\cite{Pratt_2011} Ni substitution also manifests incommensurate (IC) AFM order over a narrow range of $x$ at approximately half of the critical concentration for IC order in Co. However, the AFM ordering for Cu substitution remains commensurate (C) up to $x \approx 0.044$, where AFM order is absent, demonstrating that a rigid-band view is not appropriate.  We propose that the absence of incommensurability and superconductivity for Ba(Fe$_{1-x}$Cu$_x$)$_2$As$_2$ arises from enhanced impurity scattering associated with Cu, consistent with the behavior of $T_c$ with Cu substitution in Ba(Fe$_{1-x-y}$Co$_x$Cu$_y$)$_2$As$_2$.

{\par}Single crystals of Ba(Fe$_{1-x}M_x$)$_2$As$_2$ with $M$ = Co, Ni and Cu were grown out of a FeAs self-flux using the high temperature solution growth technique described in Ref.~\onlinecite{canfield_2009,ni_cu_2010}. Using wavelength-dispersive spectroscopy, the combined statistical and systematic error on the $M$ composition is not greater than 5\%. Inelastic neutron scattering experiments were performed on the HB3 spectrometer at the High-Flux Isotope Reactor at Oak Ridge National Laboratory at a fixed final energy of 14.7 meV.  The data here are described in terms of the orthorhombic indexing, $\textbf{Q}$ = ($\frac{2\pi\emph{H}}{a}$ $\frac{2\pi\emph{K}}{b}$ $\frac{2\pi\emph{L}}{c}$), where $a \geq b \approx 5.6$ {\AA} and $c \approx 13$ {\AA}. Samples were aligned in the orthorhombic ($H 0 L$) plane and mounted in a closed-cycle refrigerator for low-temperature studies.  Diffraction measurements were done on the TRIAX triple-axis spectrometer at the University of Missouri Research Reactor employing an incident neutron energy of $14.7~m$eV. Samples were studied in the vicinity of $\textbf{Q}_\mathrm{AFM}$ = (1 0 3) in the ($\zeta$ $K$ 3$\zeta$) plane, allowing a search for incommensurability along the \textbf{b} axis ([0 $K$ 0], transverse direction) as found for Ba(Fe$_{1-x}$Co$_x$)$_2$As$_2$.\cite{Pratt_2011}

{\par}The inelastic neutron scattering spectra were measured for single-crystals of underdoped Ba(Fe$_{0.953}$Co$_{0.047}$)$_2$As$_2$ and Ba(Fe$_{0.972}$Cu$_{0.028}$)$_2$As$_2$.  These two samples were chosen because they have similar tetragonal-orthorhombic transition temperatures [$T_S$(Co) = 63~K, $T_S$(Cu) = 73~K] and AFM transition temperatures [$T_N$(Co) = 47~K, $T_N$(Cu) = 64~K], which are comparably suppressed relative to the parent BaFe$_2$As$_2$ compound [$T_N~\approx~T_S = 140$~K].  Bulk transport measurements show a superconducting transition for the Co substituted sample at 17 K, whereas no superconducting transition is observed in the Cu substituted sample down to 2~K.  The Co(Cu) sample consisted of 9(2) co-aligned crystals weighing a total of 1.88(1.52) grams and a total mosaic width of 1.5$^\circ$(0.6$^\circ$) full-width-at-half-maximum (FWHM).

%%% Figure 1
\begin{figure}[t!]
\centering\includegraphics[width=0.9\linewidth]{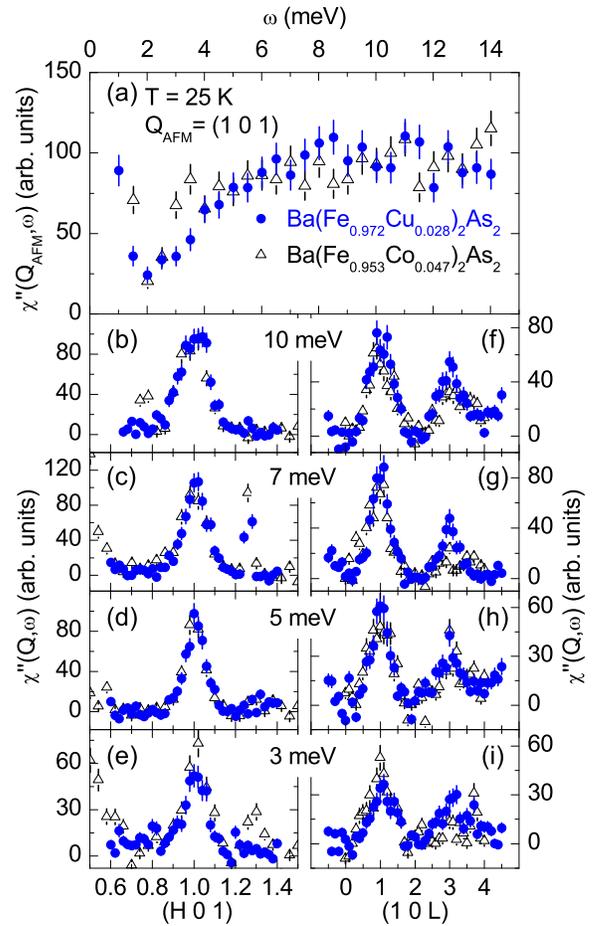}\\
\caption{(Color online) Inelastic neutron scattering from the Co and Cu substituted samples. (a) Normalized dynamic magnetic susceptibility at 25~K determined from constant-\textbf{Q} \emph{E}-scans at the (1 0 1) magnetic Bragg point. (b)-(i) \textbf{Q}-scans at several fixed values of the energy loss. The feature at (1.3 0 1) results from spurious scattering not related to spin excitations.}\label{INS}
\end{figure}

{\par}Figure~\ref{INS} compares the inelastic magnetic scattering from the Co and Cu substituted samples measured at $T$ = 25~K.  The data are plotted in terms of the dynamic magnetic susceptibility, $\chi''(\textbf{Q},\omega) = [I(\textbf{Q},\omega)-C(\textbf{Q},\omega)](1-e^{-\hbar\omega/kT})$, where $I(Q,\omega)$ is the raw neutron intensity and $C(\textbf{Q},\omega)$ is the nonmagnetic background determined from averaged inelastic scattering at positions well away from the magnetic signal [e.g. \textbf{Q} = (0.79 0 1.72) and (0.72 0 1.88)].  The data for these samples were normalized to each other using measurements of several transverse phonon peaks, and this was found to be consistent with the ratio of the masses of the two samples.

{\par}The constant-$\textbf{Q}$ energy scan measured at \textbf{Q$_\mathrm{AFM}$} = (1 0 1) [Fig.~\ref{INS}(a)] as well as the constant-$E$ $\textbf{Q}$-scans along the [1 0 0] and [0 0 1] directions [Figs.~\ref{INS}(b)-(i)] show that the normal state (above $T_c$) dynamic susceptibility for Co and Cu substituted samples are indistinguishable.  Below $T_c$, the spectrum of the Co substituted sample manifests a magnetic resonance feature above 4 meV (not shown here) in the superconducting state as observed previously by many groups.\cite{Johnston_2010,LandC_2010}  However, the dynamic susceptibility for the nonsuperconducting Cu substituted sample is temperature independent down to 5 K.  The close similarity of the normal state susceptibility for single substitutions of Co and Cu show that we must look beyond the spin fluctuation spectra to understand the absence of superconductivity in Ba(Fe$_{1-x}$Cu$_x$)$_2$As$_2$, motivating a closer look at the effects of $M$ substitutions upon magnetism in BaFe$_2$As$_2$ .

%%% Figure 2
\begin{figure}[t!]
\centering\includegraphics[width=0.9\linewidth]{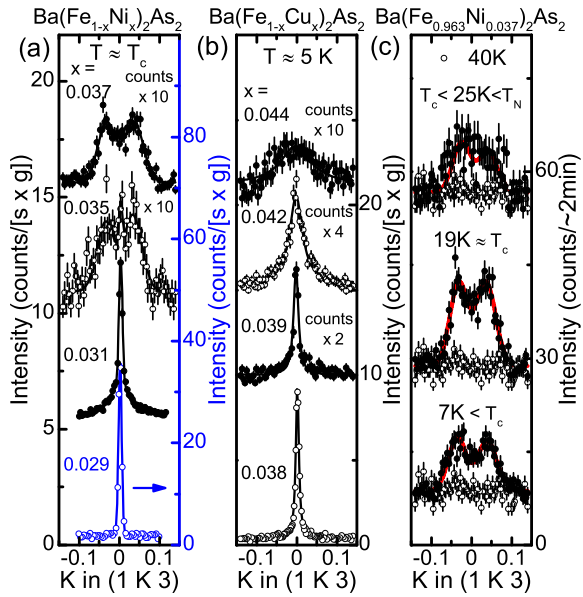}\\
\caption{(Color online) Scattering near the (1 0 3) magnetic Bragg point for Ba(Fe$_{1-x}M_x$)$_2$As$_2$ where $M$ is (a) Ni and (b) Cu. (c) Temperature dependence of the scattering near the (1 0 3) magnetic Bragg point for Ba(Fe$_{0.963}$Ni$_{0.037}$)$_2$As$_2$. Intensities are normalized by  mass of the samples to facilitate comparisons. Lines are fits to the data, as described in the text.}\label{NiCu1}
\end{figure}

{\par}We have shown previously that IC-AFM order is found in Ba(Fe$_{1-x}$Co$_x$)$_2$As$_2$ for $x~\geq~0.056$, providing a measure of the effect of Co substitution on the Fermi surface. Co substitution detunes the electron- and hole-like Fermi surfaces\cite{CLiu_2011} and eventually results in a mismatch that favors IC-AFM order. This suggests that Fermi surface nesting is a crucial factor in stabilizing both C and IC phases in the magnetic phase diagram of the $A$Fe$_2$As$_2$ ($A$ = Ba, Sr, Ca) compounds.\cite{Pratt_2011}

Figures~\ref{NiCu1}(a) and (b) show the low-$T$ scattering for transverse (0 $K$ 0) scans through the (1 0 3) magnetic Bragg point for several Ni and Cu compositions.  For Ba(Fe$_{1-x}$Ni$_x$)$_2$As$_2$, a transition from a C-AFM order for $x < 0.035$ (with resolution limited magnetic Bragg peaks) to IC-AFM order for $x \geq 0.035$ is clearly demonstrated by the symmetric pair of peaks at (1 $\pm\epsilon$ 3).  For $x > 0.037$, no long-range AFM order was observed. The lines in Fig.~\ref{NiCu1}(a) are fits to the data using a single Gaussian for $x = 0.029$, a convoluted Gaussian + Lorentzian line shape for $x = 0.031$, three Gaussians for $x = 0.035$ (to account for the presence of the dominant IC and residual C components), and two Gaussians for $x = 0.037$. The detailed description of the IC structure based on these fits is very similar for Co and Ni substitution. The incommensurability, $\epsilon$, derived from fits to these data was $0.033\pm0.003$ reciprocal lattice units (r.l.u.), close to the value found for $\epsilon$ for Co  samples.\cite{Pratt_2011}

{\par}These data show that, as previously observed for Co substitution, Ni substitution results in an abrupt change from C to IC AFM order at $x_c = 0.035 \pm 0.002$. The ratio ($\approx 0.6$) of this critical concentration to $x_c = 0.056$ for Co\cite{Pratt_2011}, is consistent with Ni "donating" roughly twice the number of electrons as Co. As discussed previously for Co substitutions, the abrupt transition between C and IC magnetic structures is similar to what has been observed for dilute substitutions of Mn or Ru in the canonical spin-density-wave (SDW) system, Cr.\cite{Pratt_2011}  Detailed theoretical studies of the nesting and free energy of the competing C and IC-SDW states in BaFe$_2$As$_2$ may shed further light on this behavior.

\begin{figure}[t!]
\centering\includegraphics[width=0.9\linewidth]{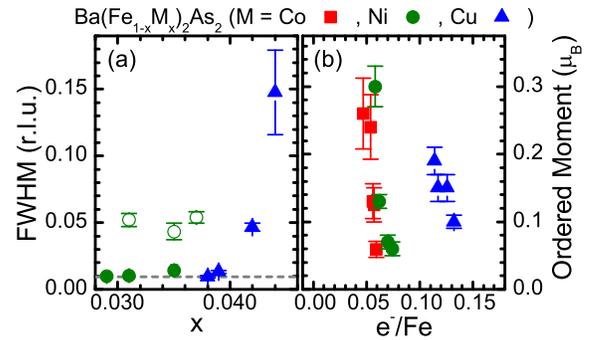}\\
\caption{(Color online)  Trends in the FWHM and maximum ordered moment for $M$ substitution. (a) Evolution of the FWHM of the magnetic peaks vs. concentration. The solid(open) circles represent the FWHM of the C-AFM(IC-AFM) peaks. (b) Measured ordered moment derived from the integrated intensity of the magnetic Bragg peaks as a function of the extra electron count, assuming that Co donates 1, Ni 2, and Cu 3, extra-electrons to the $d$-band.  The data for Ba(Fe$_{1-x}$Co$_x$)$_2$As$_2$ are taken from references~\onlinecite{Pratt_2011,fernandes_unconventional_2010}.} \label{MnW1}
\end{figure}

{\par}There is a significant broadening of the IC magnetic diffraction peaks as compared to the C magnetic peaks indicating a much reduced magnetic correlation length ($\xi \sim$ 60~{\AA}), again consistent with the broadening found for the Co substituted samples.\cite{Pratt_2011} The peak widths obtained from these fits are given in Fig.~\ref{MnW1}(a) and show that the C component remains resolution limited, whereas the IC peaks are more than $5$ times broader.  Recent measurements on Ni-substituted samples by Luo \emph{et al.}\cite{Luo_2012} are consistent with our results. The temperature dependence of the transverse (0 $K$ 0) scans through the magnetic scattering for superconducting Ba(Fe$_{0.963}$Ni$_{0.037}$)$_2$As$_2$ is illustrated in Fig.~\ref{NiCu1}(c).  The integrated intensity of the magnetic scattering increases below $T_N$, reaches a maximum at the superconducting transition temperature ($T_c$), and decreases monotonically below $T_c$ as observed previously for Co substituted samples,\cite{pratt_coexistence_2009,christianson_2009,fernandes_unconventional_2010,Pratt_2011} demonstrating, again, that magnetic order competes with superconductivity.  The positions and widths of the IC magnetic peaks appear to be temperature independent within the resolution of our measurement.

%%% Figure 3

{\par}In striking contrast to the data for Co samples\cite{Pratt_2011} and here for Ni, Figure~\ref{NiCu1}(b) shows no evidence of a C-to-IC transition versus $x$ for Ba(Fe$_{1-x}$Cu$_x$)$_2$As$_2$.  Instead, the C magnetic Bragg peak is well described by a single Lorentzian lineshape that broadens strongly for $x \geq 0.039$ [see Figs.~\ref{NiCu1}(b) and \ref{MnW1}(a)], and no AFM long-range order is found for $x \geq 0.044$.  To further emphasize the differences between Co, Ni and Cu substitutions, Fig.~\ref{MnW1}(b) displays the maximum ordered magnetic moment (at $T_c$ for Co and Ni substitution and at our base temperature, 5~K for Cu substitution) as a function of extra electron count under the oft-used assumption that Co, Ni, and Cu donate 1, 2, and 3, respectively, to the $d$-bands. The maximum ordered moment was estimated from the integrated intensity of the magnetic Bragg peaks  using the C magnetic structure factor normalized by the mass of the samples, as described previously.\cite{fernandes_unconventional_2010}  Under the stated assumption, Co and Ni act similarly to suppress the moment over a range of $x$ that mimics a rigid-band picture.  This is clearly not the case for Cu substitution (although rescaling the electron count by an additional factor of two would move the results on top of Co and Ni).  Nevertheless, the IC magnetic order found for Ni and Co substitutions in this regime is not found for Cu substitution.

%%% Figure 4
\begin{figure}[t!]
\centering\includegraphics[width=.9\linewidth]{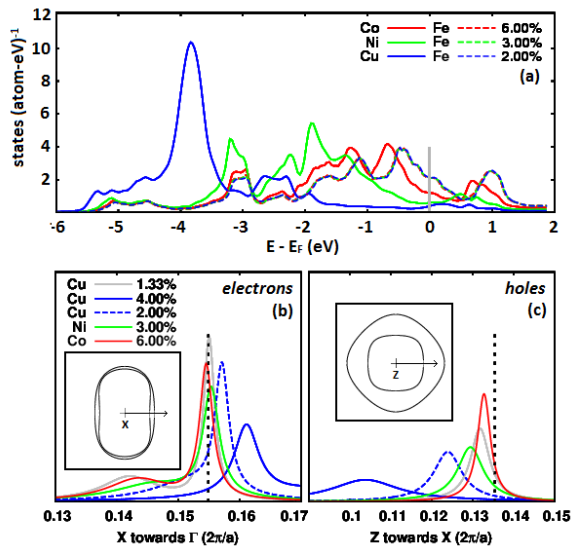}\\
\caption{(Color online) For Ba(Fe$_{1-x}M_x)$$_2$As$_2$, the KKR-CPA (a) site-projected DOS versus $E - E_F$ at  6\% Co, 3\% Ni (fixed $e^{-}/Fe$),  and 2\% Cu (the Fe DOS changes negligibly with $M$); and Bloch spectral functions, $A({\bf k};E_{F})$, along specific ${\bf k}$-directions versus at. \% $M$ for (b) electrons, and (c) holes. Insets: ${\bf k}$-direction of the cut across \emph{electron} (centered at $X$) and \emph{hole} (centered at $Z$) states.
Peak locations of electron/hole states are compared to the "rigid-band" expectations  (vertical dashed lines) from parent compound at fixed $e^-/Fe$ and three at. \% Cu values.} \label{DOS}
\end{figure}

{\par}To further elucidate the differences between Co/Ni and Cu substitution in BaFe$_2$As$_2$ we employed the Korringa-Kohn-Rostoker method using the Coherent-Potential Approximation (KKR-CPA) to address the effects of substitution on the density of states (DOS), and solute disorder (impurity) scattering on the Fermi surfaces [i.e., the Bloch spectral functions $A({\bf k};E_{F})$ at the Fermi energy $E_F$].\cite{Johnson,Alam2009,Alam2010} First, Figure~\ref{DOS}(a) shows that the $d$-band partial DOS of Co and Ni are common-band-like (e.g. overlap with the Fe $d$-bands), whereas Cu exhibits split-band behavior with its $d$-states located $\sim$4~eV below $E_F$. Only $s$-$p$ states participate at $E_F$ and, therefore, Cu behaves almost as a +1 $s$-$p$ valence with very different scattering behavior from Co and Ni. We note that these results are consistent with ordered DFT calculations at large $x$.\cite{Wadati_2010}

{\par}For nesting-driven ordering,\cite{Gyorffy_1983,Staunton_1990,Althoff_1995,Clark_1995} the convolution between the electron- and hole-like Fermi surfaces dictate the location of peaks in the susceptibility.\cite{Gyorffy_1983}  Figures~\ref{DOS}(b) and (c) illustrate the behavior of the Fermi-surfaces for \emph{electrons} and \emph{holes} at a fixed solute concentrations for Co (6\%) and Ni (3\%) [red and green lines] compared to the "rigid-band" expectation from the parent compound at a fixed $e^{-}/Fe$ (0.06). These solute concentrations are close to the respective $x_c$ for the observed C to IC magnetic ordering, and a rigid-band treatment [vertical dashed lines in Figs.~\ref{DOS}(b) and (c)] provides an estimate for $\varepsilon$ of $\sim 0.021$, similar to that observed in our measurements.  As solute content increases, the electron(hole) surfaces expand(contract) and the spectral broadening due to chemical disorder is evident. Due to common $d$-band behavior for Co and Ni, spectral peaks for the electrons clearly mimic rigid-band expectations at fixed $e^{-}/Fe$, but the holes less so.

In contrast, with a split Cu $d$-band, rigid-band concepts are rendered invalid. States well below $E_F$ (due to hybridization and band-filling) and at $E_F$ contribute to the total susceptibility.\cite{Pinski_1991,Staunton_1994,Althoff_1995,Staunton_1999} As a stronger scatterer than Fe, Co, or Ni, $\sim$1\% Cu (rather than 2\% Cu assuming a +3 valence) acts like 6\% Co or 3\% Ni in terms of broadening of the spectral features.  Most importantly, the hole states are especially sensitive to the Cu content, with a rapid loss of intensity and increased disorder broadening evident, as shown for up to 4\% Cu for comparison with our experiments.   The convolution of the electron- and hole-like Fermi surfaces is dramatically diminished and, therefore, so is the impetus for incommensurability.\cite{Althoff_1995}

{\par}We propose that the absence of IC-AFM order in Ba(Fe$_{1-x}$Cu$_x$)$_2$As$_2$ arises from enhanced impurity scattering effects associated with the stronger potential for Cu.  The small incommensurability measured for Co and Ni substituted BaFe$_2$As$_2$ requires relatively sharp and well-defined features in the Fermi surface topology.  Disorder due to impurity scattering introduces spectral broadening in both energy and momentum to the extent that the magnetic structure remains C rather than IC for Cu.  This is in substantial agreement with recent work by Berlijn \emph{et al.,}\cite{Berlijn_2011} for Zn substitutions in BaFe$_2$As$_2$.

{\par}Finally, we note that such impurity effects are expected to impact superconductivity in the iron pnictides as well. Essential elements of the under-doped regions of the phase diagram for electron-doped BaFe$_2$As$_2$ are captured by considering both inter- and intra-band impurity scattering.\cite{Vavilov_2011,fernandes_2012}  Although impurity scattering introduced by $M$ substitution causes pair breaking and suppresses $T_c$, it can be even more damaging for spin-density wave ordering so that $T_N$ is suppressed more rapidly, allowing superconductivity to emerge at finite substitution levels. Interestingly, the phenomenological model by Fernandes \emph{et al.}\cite{fernandes_2012} indicates that the behavior of $T_c$ for s$^{\pm}$ pairing is a non-monotonic function of impurity concentration, depending on the strength of the impurity potential and the ratio of the intra-band ($\Gamma_0$) to inter-band ($\Gamma_\pi$) impurity scattering, which may vary strongly between Co and Cu.  Indeed, they find a range in $\frac{\Gamma_0}{\Gamma_\pi}$ where $T_c$ first increases and then decreases with impurity concentration, very similar to that observed for Co/Cu co-substitutions in BaFe$_2$As$_2$.

\begin{acknowledgments}
This work was supported by the U.S. Department of Energy (DOE), Office of Basic Energy Sciences (OBES), Division of Materials Sciences and Engineering. Work at the High Flux Isotope Reactor, Oak Ridge National Laboratory, was sponsored by the Scientific User Facilities Division, DOE/OBES. SNK and DDJ acknowledge partial support from ORNL's Center for Defect Physics, Energy Frontier Research Center.
\end{acknowledgments}

\bibliographystyle{apsrev}
\bibliography{TMdoping}

\end{document}